\documentclass[12pt]{article}
\usepackage{graphicx}

\newcommand{\be}{\begin{equation}}
\newcommand{\ee}{\end{equation}}
\newcommand{\bea}{\begin{eqnarray}}
\newcommand{\eea}{\end{eqnarray}}
\newcommand{\nn}{\nonumber}

\newcommand\pubnumber{BARI-TH/2014-693}
\newcommand\pubdate{\today}

\def\bari{Istituto Nazionale di Fisica Nucleare - Sezione di Bari\\Via Orabona 4, I-70126 Bari, ITALY}

\def\Title#1{\begin{center} {\Large #1 } \end{center}}
\def\Author#1{\begin{center}{ \sc #1} \end{center}}
\def\Address#1{\begin{center}{ \it #1} \end{center}}

\newcommand\pubblock{\rightline{\begin{tabular}{l} \pubnumber\\
         \pubdate  \end{tabular}}}
\newenvironment{Abstract}{\begin{quotation}  }{\end{quotation}}
\newenvironment{Presented}{\begin{quotation} \begin{center} 
             PRESENTED AT\end{center}\bigskip 
      \begin{center}\begin{large}}{\end{large}\end{center} \end{quotation}}
\def\Acknowledgements{\bigskip  \bigskip \begin{center} \begin{large}
             \bf ACKNOWLEDGEMENTS \end{large}\end{center}}

\def\beq{\begin{equation}}
\def\eeq#1{\label{#1}\end{equation}}
\def\eeqn{\end{equation}}

\def\beqa{\begin{eqnarray}}
\def\eeqa#1{\label{#1}\end{eqnarray}}
\def\eeqan{\end{eqnarray}}

\let\bar=\overbar

\def\Dslash{\not{\hbox{\kern-4pt $D$}}}
\def\dslash{\not{\hbox{\kern-2pt $\del$}}}

\def\ee{e^+e^-}

\def\msb{{\bar{\ssstyle M \kern -1pt S}}}

\begin{document}
\begin{titlepage}
\pubblock

\vfill
\Title{New Physics Scenarios in $b \to c \ell {\bar \nu}_\ell$ decays}
\vfill
\Author{ Fulvia De Fazio}
\Address{\bari}
\vfill
\begin{Abstract}
\noindent The latest BaBar measurements of the ratios ${\cal R}(D^{(*)})=\displaystyle{\frac{{\cal B}(B \to D^{(*)} \tau {\bar \nu}_\tau)}{{\cal B}(B \to D^{(*)} \mu {\bar \nu}_\mu)}}$  deviate from the Standard Model predictions at the global level of 3.4$\sigma$.  
A possibility to reproduce these experimental ratios  without affecting other modes which do not show similar deviations is to 
 consider new physics scenarios producing an
additional tensor operator in the effective weak Hamiltonian. I describe the impact of such an operator  in semileptonic $B \to D^{(*)}$ modes and  in semileptonic $B$ and $B_s$ decays to excited positive  parity charmed  mesons. In particular, I discuss the most effective observables able to discriminate new physics from  the Standard Model.
\end{Abstract}
\vfill
\begin{Presented}
8th International Workshop on the CKM Unitarity Triangle (CKM 2014), \\Vienna, Austria, September 8-12, 2014
\end{Presented}
\vfill
\end{titlepage}
\def\thefootnote{\fnsymbol{footnote}}
\setcounter{footnote}{0}

\section{Introduction}
Semileptonic decays induced by the $b \to c \ell {\bar \nu}_\ell$ transition are the cleanest modes to measure the element $V_{cb}$ of the Cabibbo-Kobayashi-Maskawa (CKM) matrix. Recently, the possibility to use them to test lepton flavour universality and to reveal new physics (NP) effects  emerged, prompted by the  BaBar Collaboration results \cite{Lees:2012xj}:
\bea
{\cal R}^-(D)&=&\frac{{\cal B}(B^- \to D^0 \tau^- \,{\bar \nu}_\tau)}{{\cal B}(B^- \to D^0 \ell^- \,{\bar \nu}_\ell)}=0.429 \pm 0.082 \pm 0.052  \,\,\, , \nn \\
{\cal R}^-(D^*)&=&\frac{{\cal B}(B^- \to D^{*0} \tau^- \,{\bar \nu}_\tau)}{{\cal B}(B^- \to D^{*0} \ell^- \, {\bar \nu}_\ell)}=0.322 \pm 0.032 \pm 0.022\,\, , \nn \\
{\cal R}^0(D)&=&\frac{{\cal B}({\bar B}^0 \to D^+ \tau^- \, {\bar \nu}_\tau)}{{\cal B}({\bar B}^0 \to D^+ \ell^- \, {\bar \nu}_\ell)}=0.469 \pm 0.084 \pm 0.053 \,\,\, ,\nn \\
{\cal R}^0(D^*)&=&\frac{{\cal B}({\bar B}^0 \to D^{*+} \tau^- \, {\bar \nu}_\tau)}{{\cal B}({\bar B}^0 \to D^{*+} \ell^- \, {\bar \nu}_\ell)}=0.355 \pm 0.039 \pm 0.021 \,\,  \label{data}
\eea
(the first and second error are the statistic and systematic uncertainty, respectively). As I discuss in the following, the results in (\ref{data})  globally  deviate at  3.4$\sigma$ level with respect to the Standard Model (SM)  predictions   \cite{Lees:2012xj,Fajfer:2012vx}, 
and they might be due to  new particles with large couplings to the heavier fermions, namely charged scalars   contributing to  tree-level $b \to c \ell \bar \nu$ transitions
\cite{Fajfer:2012vx,altri}.
However, if such new particles exist, they should  also affect the purely  leptonic  $B^- \to \tau^- {\bar \nu}_\tau$ mode, for which the most recent experimental branching ratio determinations are compatible with the  SM prediction  \cite{Adachi:2012mm}.
To assess whether the results  in (\ref{data}) are a signal of NP, one should investigate which NP scenario can reproduce them without affecting the  leptonic mode. Here, I summarize the analysis  in \cite{Biancofiore:2013ki} devoted to such an issue.

\section{Exclusive $b \to c \ell {\bar \nu}_\ell$ decays}
A  modification of the SM effective weak Hamiltonian  that produces a variation of  the ratios (\ref{data}),   leaving the purely  leptonic $B$ decays unaffected, is
\be
H_{eff}= {G_F \over \sqrt{2}}V_{cb} \left[ {\bar c} \gamma_\mu (1-\gamma_5) b \, {\bar \ell} \gamma^\mu (1-\gamma_5) {\bar \nu}_\ell + \epsilon_T^\ell \, {\bar c} \sigma_{\mu \nu} (1-\gamma_5) b \, {\bar \ell} \sigma^{\mu \nu} (1-\gamma_5) {\bar \nu}_\ell \right] \,\,\, . \label{heff}
\end{equation}
$G_F$ is the Fermi constant, and  
a new tensor operator has been introduced, with the coupling  $\epsilon_T^\ell$  assumed to mainly contribute for  $\ell=\tau$:
  $\epsilon_T^{e,\mu}=0$ and  $\epsilon_T\equiv \epsilon_T^\tau$. 
 Physical observables allow us  to  constrain this coupling.

To compute the branching ratios  in   (\ref{data}), the hadronic matrix  element of the  Hamiltonian (\ref{heff}) between the  $B$ and $D^{(*)}$ mesons are required. In the case of  $B \to D \ell {\bar \nu}_\ell$,  they can be parameterized in terms of  the form factors $F_0(q^2),\,F_1(q^2)$ for the SM part, and  of $F_T(q^2)$, $G_T(q^2)$ for the tensor operator.   In the case of $B \to D^* \ell {\bar \nu}_\ell$, the form factors are   $A_1(q^2)$,$A_2(q^2)$, $A_0(q^2)$, $V(q^2)$ and $T_0(q^2)$, $T_1(q^2)$, $T_2(q^2)$, $T_3(q^2)$, $T_4(q^2)$ and $T_5(q^2)$, respectively (see \cite{Biancofiore:2013ki} for the matrix element parametrization).
In the  infinite heavy quark mass limit, all   those form factors   can be related to the Isgur-Wise universal function $\xi$ \cite{Isgur:1989ed}. 
 In \cite{Biancofiore:2013ki}  the results in \cite{Caprini:1997mu} for such relations  have been used, which include next to leading order corrections both in  $1/m_{b,c}$ and in ${\cal O}(\alpha_s)$. Moreover, 
 the experimental results of the
 BaBar  analysis of   $B \to D \mu {\bar \nu}_\mu$ \cite{Aubert:2009ac} and of Belle analysis of $B \to D^* \mu {\bar \nu}_\mu$  \cite{Dungel:2010uk} have been exploited to fix the parameters of the  function $\xi$. 
As a result, the SM predictions for the ratios in  (\ref{data})  are:
$ {\cal R}^0(D)\Big|_{SM}=0.324 \pm 0.022 $ and
$ {\cal R}^0(D^*)\Big|_{SM}=0.250 \pm 0.003$.
While the data for $ {\cal R}^{-,0}(D)$ do not deviate significantly from  SM, a  discrepancy is found in  $ {\cal R}^{-,0}(D^*)$.
In the next section I describe the impact of the new Hamiltonian   (\ref{heff}) on these ratios.

\section{NP predictions for ${\cal R}(D^{(*)})$ and other observables }
\begin{figure}[!b]
\centering
\includegraphics[width = 0.37\textwidth]{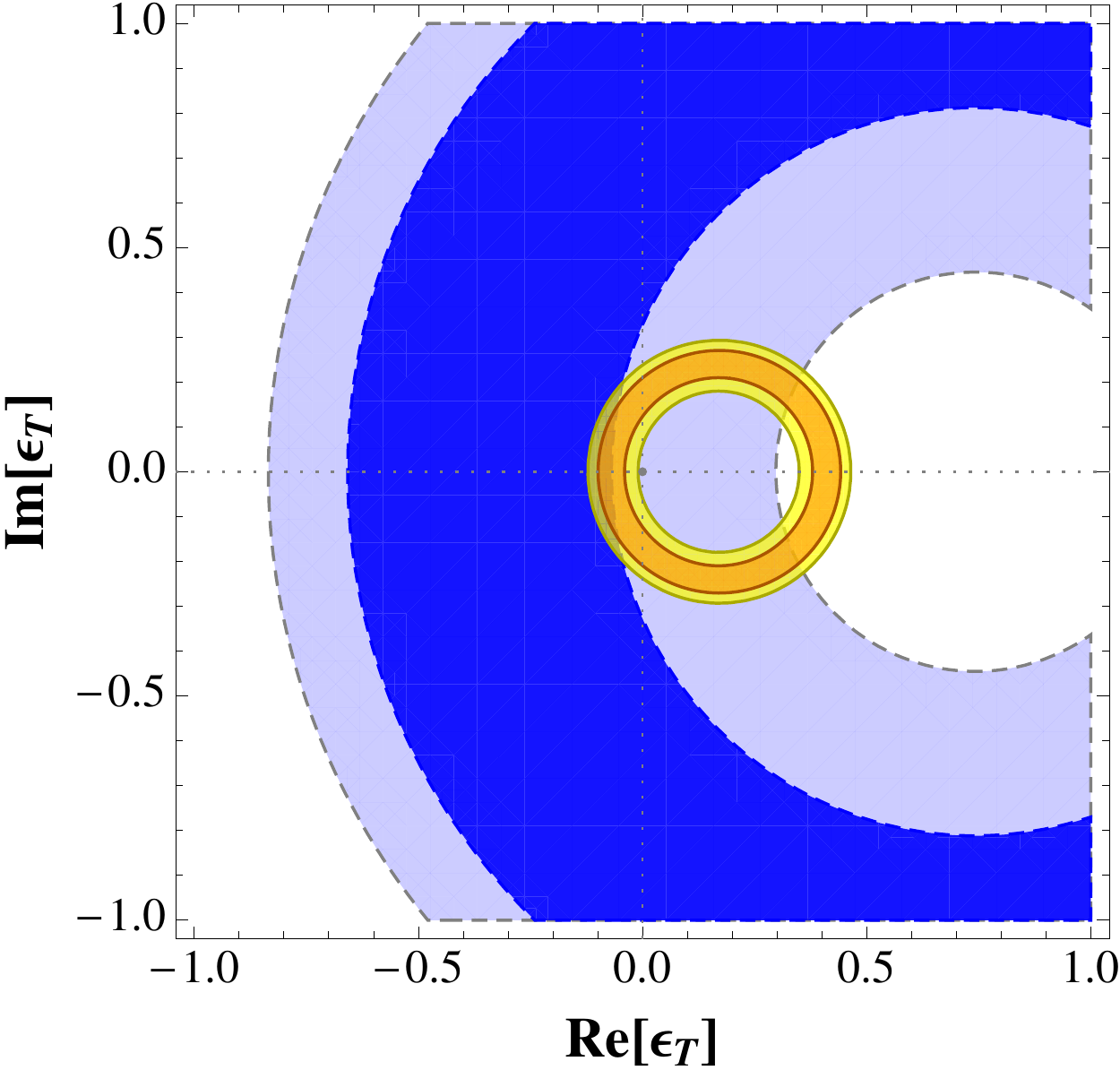} 
\caption{ Region in the plane $(Re(\epsilon_T),Im(\epsilon_T))$  determined from the experimental data  in (\ref{data})  on  ${\cal R}(D)$  (large rings) and ${\cal R}(D^*)$ (small rings).  The inner and outer circles correspond to  $1 \sigma$ and $2 \sigma$, respectively. }\label{fig:oases}
\end{figure}

The inclusion of the new tensor operator in (\ref{heff})  modifies the SM prediction for ${\cal R}(D^{(*)})$.
Therefore,  the  coupling $\epsilon_T$ can be bound  in order to reproduce the results (\ref{data}); predictions for other observables can be worked out with $\epsilon_T$ varying in the obtained region  \cite{Biancofiore:2013ki}.
The experimental constraints bound the region 
 shown in Fig.\ref{fig:oases}.  The biggest circle represents the constraint from the $ {\cal R}(D)$ data,  the smaller circle is derived imposing the result for ${\cal R}(D^*)$.
Adopting the parameterization
$\epsilon_T=|a_T| e^{i \theta}+\epsilon_{T0}$,
the result depicted in Fig.\ref{fig:oases} can be written as
\be
Re[\epsilon_{T0}]=0.17\,, \,\,\, \,\, \,\, Im[\epsilon_{T0}]=0 \,\,\, ,\,\,\, \,\, \,\,
|a_T| \in [0.24,\,0.27]\,,  \,\,\,  \,\, \,\,\theta \in [2.6,\,3.7]\, {\rm rad} \,\,. \label{ranges-for-epsilonT}
\end{equation}
 Several observables can be computed varying $\epsilon_T$ in this range \cite{Biancofiore:2013ki}. 

\begin{figure}[!b]
\centering
\includegraphics[width = 0.37\textwidth]{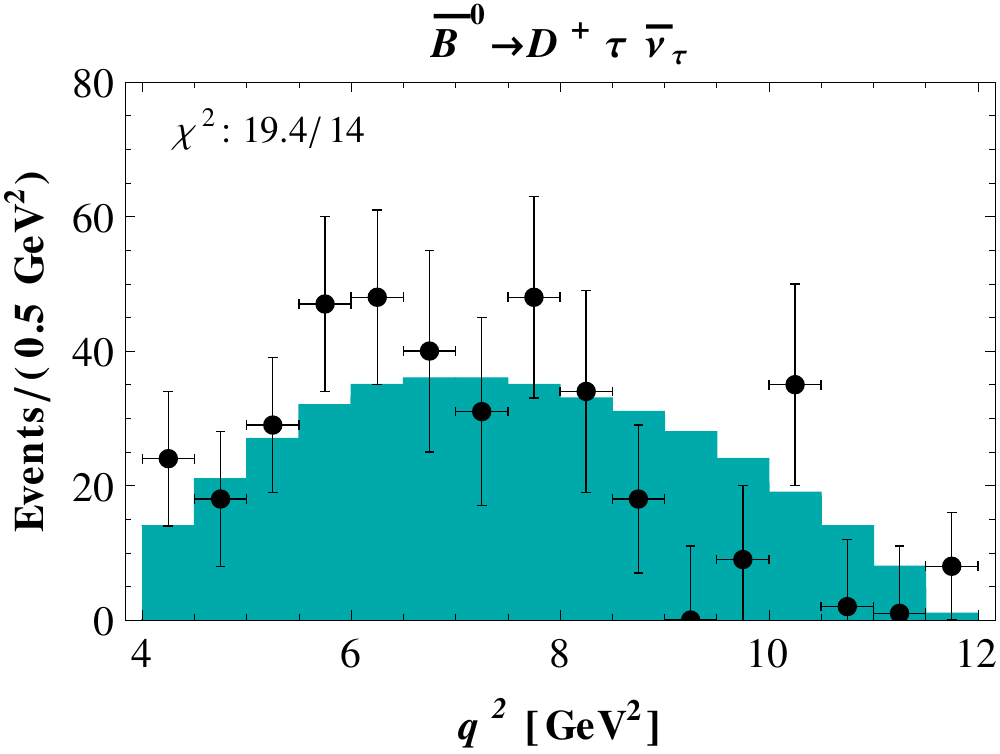}\hspace*{1cm}
\includegraphics[width = 0.37\textwidth]{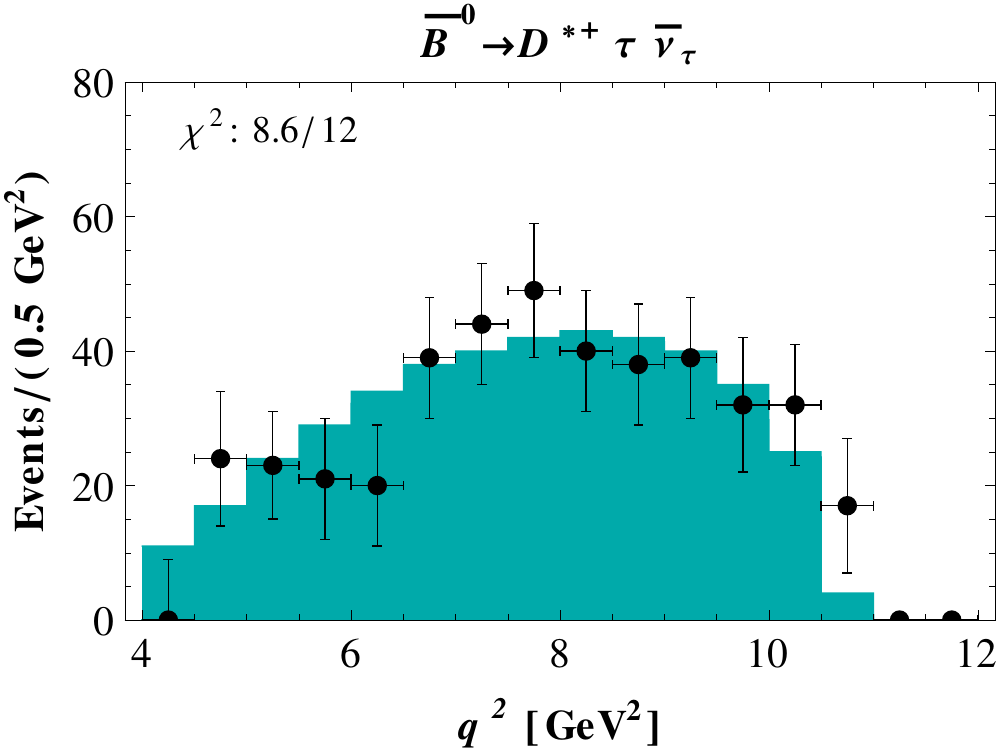}\\
\caption{
 $\displaystyle{\frac{d \Gamma(B \to D \tau {\bar \nu}_\tau)}{dq^2}}$ (left) and $\displaystyle{\frac{d \Gamma(B \to D^* \tau {\bar \nu}_\tau)}{dq^2}}$ (right) distributions in the NP scenario (for the central value of $\epsilon_T$, shaded histograms), compared to the BaBar data (points) \cite{Lees:2013uzd}. The distributions are normalized to the total number of events.}
\label{fig:newspectrum}
\end{figure}

The spectra  $d \Gamma(B \to D^{(*)} \tau {\bar \nu}_\tau)/dq^2$ (summed over the $D^*$ polarizations)
 have been measured by BaBar 
 \cite{Lees:2013uzd} and compared to SM predictions, arguing that the SM distributions are compatible with data.  
 Fig.\ref{fig:newspectrum} shows that there is compatibility also in the NP scenario discussed here, therefore
  the shape of the  decay spectra is   not effective to discriminate between the two  possibilities. 
Observables  showing more significant deviations from the SM predictions  are the $D^*$ longitudinal (L) and transverse (T) polarization fractions $F_{L,T}(q^2)=\frac{d \Gamma_{L,T}(B \to D^* \tau {\bar \nu}_\tau)}{dq^2} \times  \left(\frac{d \Gamma(B \to D^* \tau {\bar \nu}_\tau)}{dq^2}\right)^{-1} $. In  SM, $F_L(q^2)$ ranges between 0.75 at low $q^2$ and about 0.35 at high $q^2$;  in  NP, with $\epsilon_T$ in the region (\ref{ranges-for-epsilonT}), $F_L(q^2)$ varies between 0.35 and about 0.65 at low $q^2$,  converging to the SM value at high $q^2$.
While in the SM the longitudinal $F_L$ dominates  at small $q^2$,  in  NP $F_L$ and $F_T$ have similar size up to $q^2=6$ GeV$^2$.

\begin{figure}[!b]
 \centering
\includegraphics[width = 0.38\textwidth]{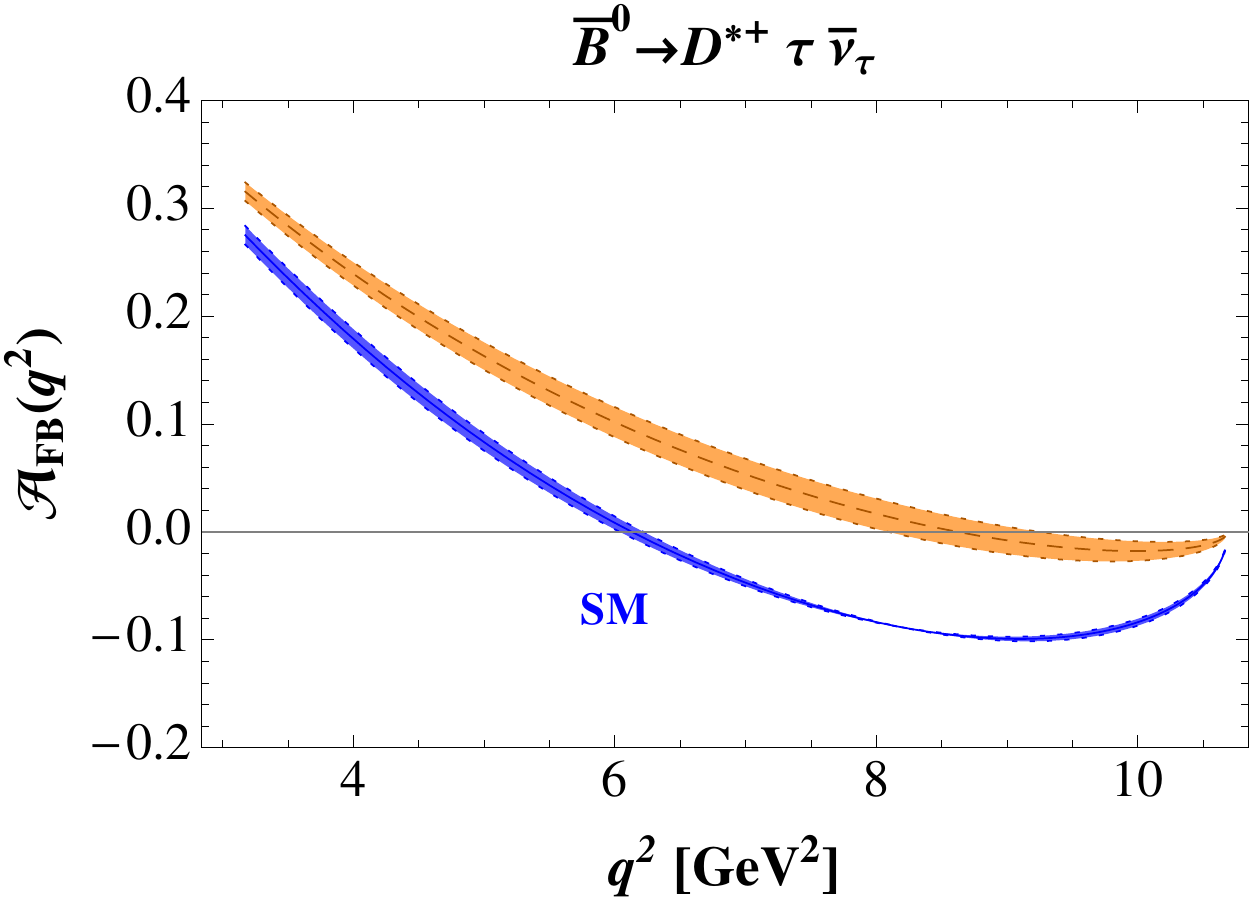}
\caption{ ${\cal A}_{FB}(q^2)$ for    $B \to D^* \tau {\bar \nu}_\tau$. The lower (blue) curve is the SM prediction,  the upper (orange) band  the NP expectation.
Uncertainty on the form factor parameters has been included and, in the case of  NP, also on the parameters $|a_T|$ and $\theta$. }\label{fig:afb}
\end{figure}

A very sensitive observable is the forward-backward  ${\cal A}_{FB}(q^2)$ asymmetry in  $ B \to D \tau {\bar \nu}_\tau$ and  $ B \to D^* \tau {\bar \nu}_\tau$. It is   defined as
\be
{\cal A}_{FB}(q^2)= \frac{\int_0^1 \,d \cos \theta_\ell \,\frac{d \Gamma}{dq^2 d \cos \theta_\ell} -\int_{-1}^0 \,d \cos \theta_\ell \, \frac{d \Gamma}{dq^2 d \cos \theta_\ell}}{\frac{d \Gamma}{dq^2}} \,\,\, ,
\label{eq:AFB}
\end{equation}
 $\theta_\ell$ being the angle between the direction of the $\tau$  and the $D^{(*)}$  in the  rest frame of the lepton pair.
Fig.\ref{fig:afb} shows ${\cal A}_{FB}(q^2)$ for     $B \to D^* \tau {\bar \nu}_\tau$.
The SM prediction  is affected by a tiny theoretical uncertainty, since the dependence on the hadronic form factors  reduces  in the  ratio.
In the case of   NP,  the prediction takes into account also the uncertainty on  $\theta$ and $|a_T|$.
The SM curve lies  below the NP distribution  for  all values of $q^2$. Interestingly,    the SM predicts a zero for ${\cal A}_{FB}$  at  $q^2\simeq 6.15$ GeV$^2$, while in the NP case the zero is shifted towards larger values:  $q^2 \in[8.1, 9.3]$ GeV$^2$.
Therefore, although the experimental determination of the zero of ${\cal A}_{FB}(q^2)$ is a challenging task,  this observable  can  discriminate  SM from  NP.

\section{Role of the tensor operator in $B \to D^{**} \ell {\bar \nu}_\ell$ decays}
\begin{figure}[!h]
 \centering
\includegraphics[width = 0.32\textwidth]{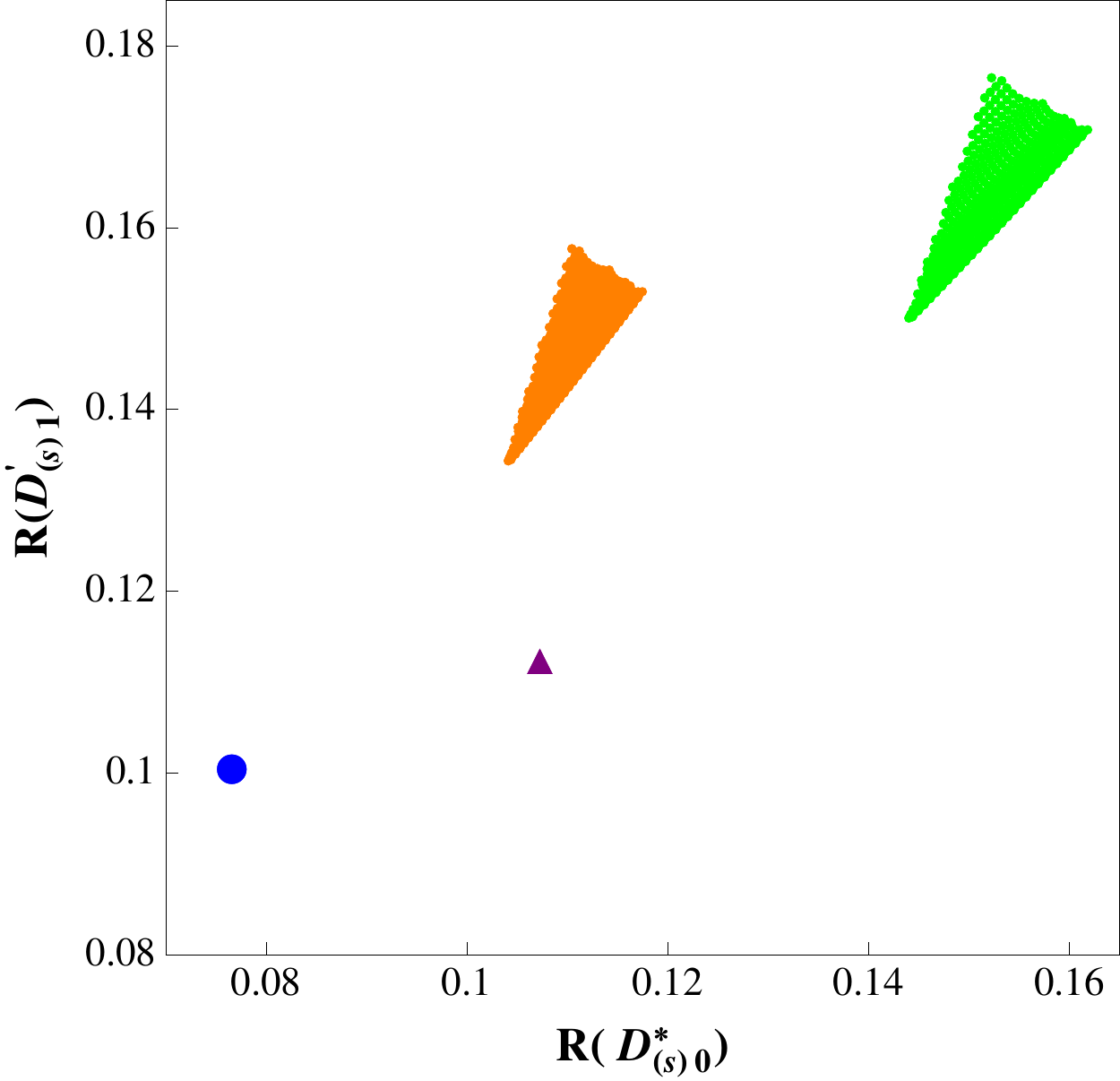}\hspace*{1.5cm}
\includegraphics[width = 0.32\textwidth]{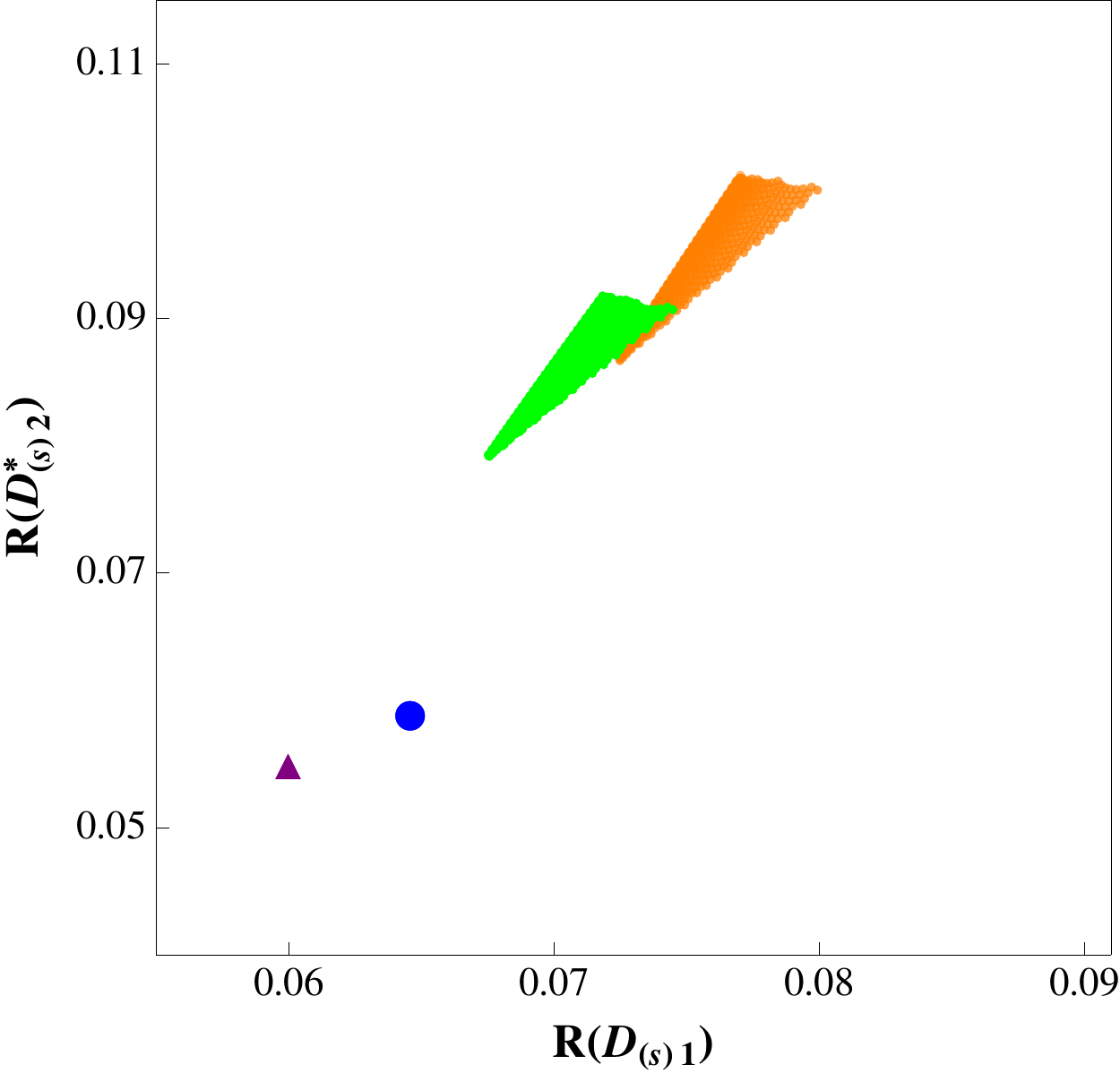}
\caption{Correlation between   ${\cal R}(D^{*}_{(s)0})$, ${\cal R}(D^\prime_{(s)1})$ (left)  and  ${\cal R}(D_{(s)1})$, ${\cal R}(D^*_{(s)2})$ (right). Orange, dark (gree, light) regions refer to mesons without (with) strangeness.  The  dots  (triangles) are the SM results for mesons  without (with) strangeness.
}\label{fig:r12-r32}
\end{figure}

\begin{figure}[!h]
 \centering
\includegraphics[width = 0.38\textwidth]{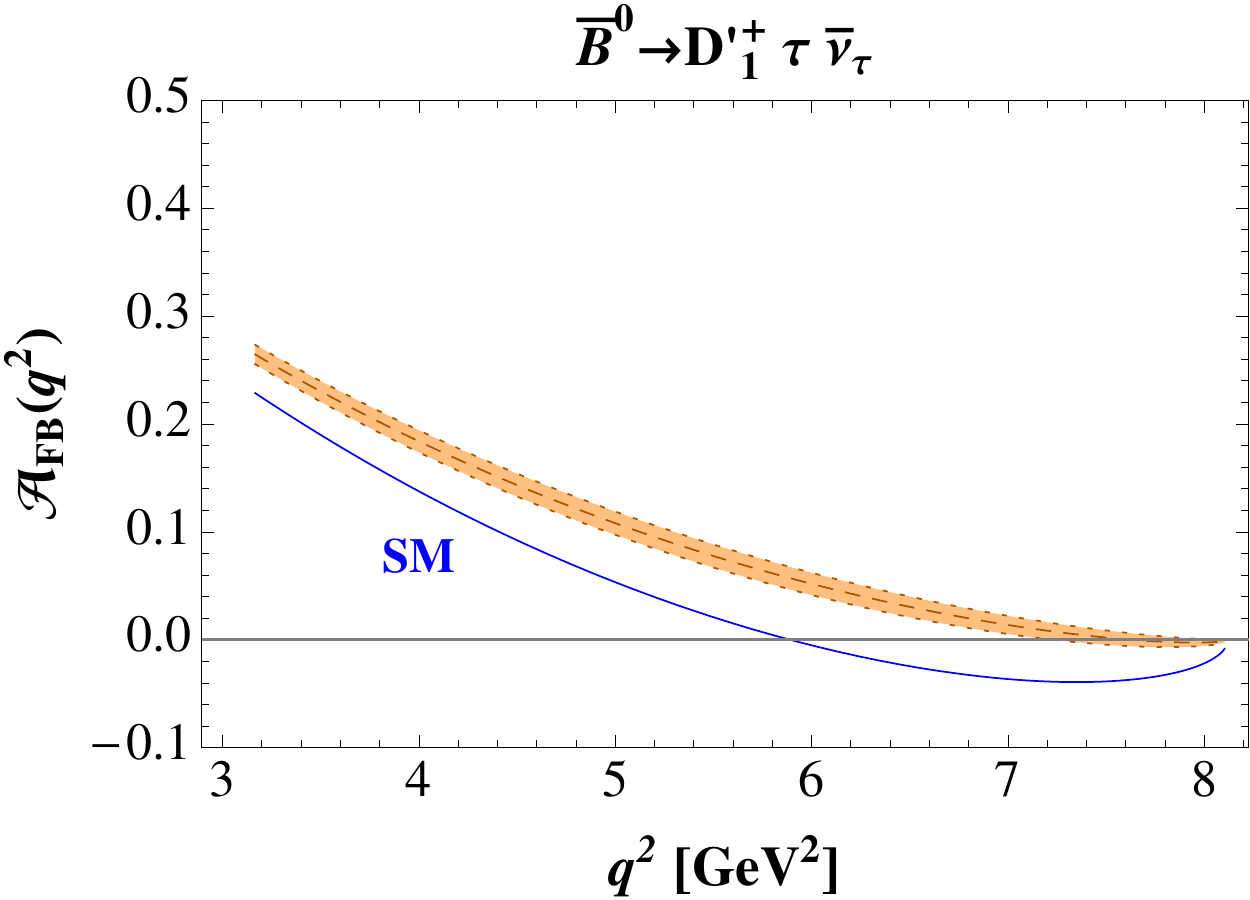}\hspace*{1cm}
\includegraphics[width = 0.38\textwidth]{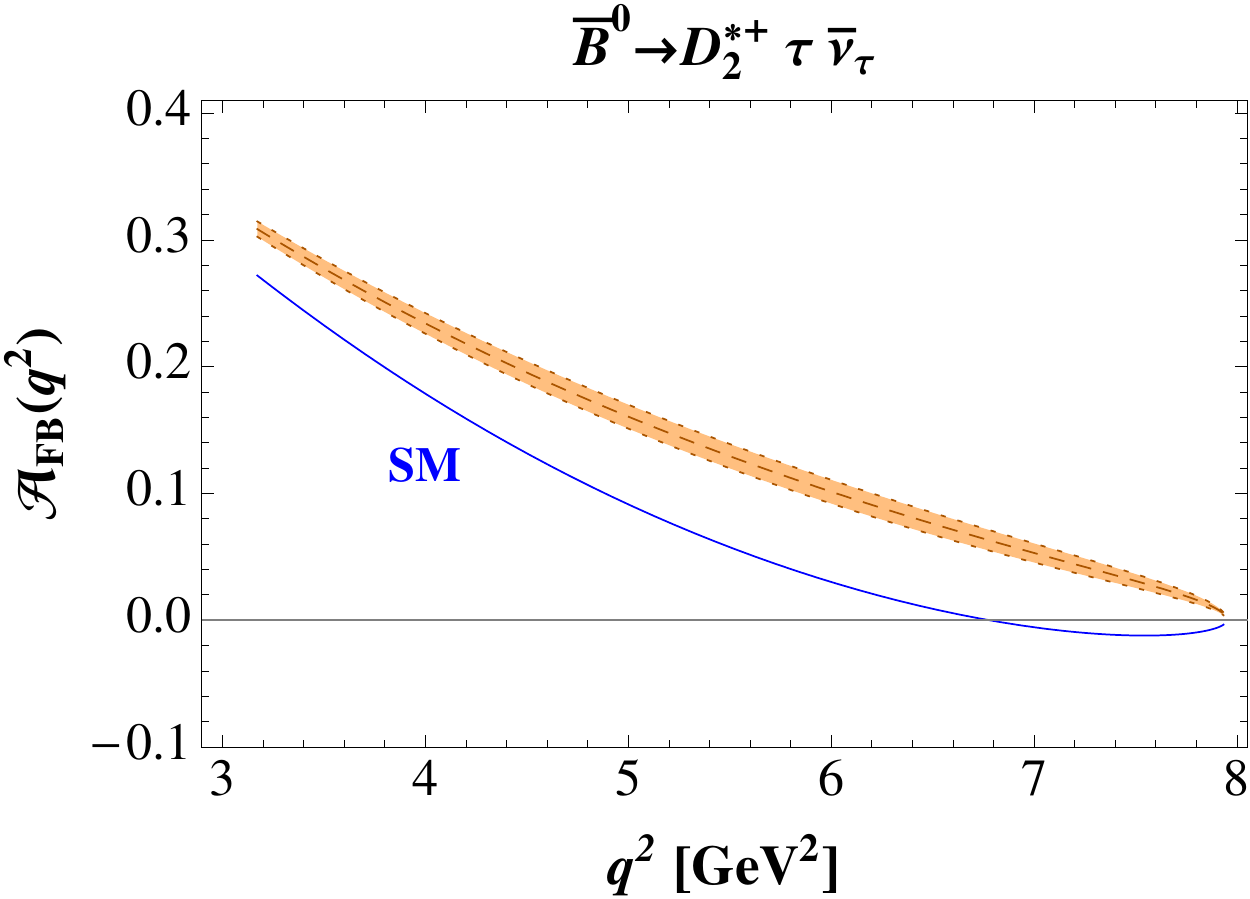}
\caption{${\cal A}_{FB}(q^2)$ for   $B \to D^\prime_1 \tau {\bar \nu}_\tau$ (left),
 and  $B \to D^*_2 \tau {\bar \nu}_\tau$  (right) . The solid (blue) curves are the SM predictions,  the dotted (orange) bands  the NP expectations. }\label{fig:afbD**}
\end{figure}

The  new operator in the effective Hamiltonian (\ref{heff}) affects  other exclusive decay modes, in particular   the semileptonic $B$ and $B_s$ transitions into  excited
charmed mesons.  In the heavy quark limit, the lightest  of such hadrons,  generically denoted as  $D_{(s)}^{**}$, can be classified in doublets:
 $(D^*_{(s)0},\,D_{(s)1}^\prime)$ have spin-parity $J^P=(0^+,1^+)$;    $(D_{(s)1},\,D_{(s)2}^*)$ have $J^P=(1^+,2^+)$. All these mesons, with and without strangeness,  have been observed with features fulfilling the theoretical expectations \cite{Colangelo:2012xi}.
The  tensor operator affects the ratios
${\cal R}(D^{**})=\displaystyle{\frac{{\cal B}(B \to D^{**} \tau \,{\bar \nu}_\tau)}{{\cal B}(B \to D^{**} \ell \,{\bar \nu}_\ell)}}$
with $D^{**}=D^*_0,\,D^\prime_1,\,D_1,\,D^*_2$, as well as the analogous ratios for
  $B_s \to D^{**}_s \ell {\bar \nu}_\ell$ transitions (the form factors are taken in the  $SU(3)_F$ symmetry limit ).
In the heavy quark limit,  the semileptonic $B$ decays to $D^{**}$ in  the same  doublet can be described in terms of a single universal function.   In the case of
$B$ decays to $(D^*_0,\,D_1^\prime)$ such a function is denoted  as $\tau_{1/2}(w)$,   for $B$ decays to $(D_1,\,D_2^*)$  by  $\tau_{3/2}(w)$ (with $q^2=m_B^2+m_{D^{**}}^2-2m_B m_{D^{**}}w$).
Several theoretical determinations of  $\tau_i(w)$ exist in the literature:
in \cite{Biancofiore:2013ki}
 a QCD sum rule
determination of $\tau_{3/2}(w)$   at leading order in  $\alpha_s$   \cite{Colangelo:1992kc}, and of  $\tau_{1/2}(w)$   at ${\cal O}(\alpha_s)$  \cite{Colangelo:1998ga} have been used.
The correlations between the ratios ${\cal R}(D^{**})$  for    $B$ and  $B_s$ decays  are displayed in Fig.\ref{fig:r12-r32}, together with
the  SM predictions: $({\cal R}(D^{*}_{0}),{\cal R}(D^\prime_{1}))=(0.077,0.100)$, $({\cal R}(D^{*}_{s0}),{\cal R}(D^\prime_{s1}))=(0.107,0.112)$, $({\cal R}(D_{1}),{\cal R}(D^*_{2})=(0.065,0.059)$ and $({\cal R}(D_{s1}),{\cal R}(D^*_{s2})=(0.060,0.055)$.
The tensor operator  produces an increase of the  ratios $\cal R$,   correlated for the two states in  each doublet.
The results are rather stable when different models for the $\tau$ functions are used, namely   those in \cite{Morenas:1997nk,Blossier:2009vy}.
Also in the case of   final states with $D^{**}$ mesons, sensitive observables are the forward-backward asymmetries. The most interesting ones are obtained in the case of  $D_1^\prime$ and
$D_2^*$,  Fig.\ref{fig:afbD**}. When the tensor operator is included,  the asymmetries ${\cal A}_{FB}$ are enhanced with respect to  SM  for all values of $q^2$. 
Moreover, the  zero of the asymmetries, that is present in both cases  in SM,  is shifted  towards larger values of $q^2$
  in  $B \to D^\prime_1 \tau {\bar \nu}_\tau$ , and  disappears  in the case of $B \to D^*_2 \tau {\bar \nu}_\tau$.
\section{Conclusions}
A possibility to explain the ratios in (\ref{data}),  leaving $B \to \tau {\bar \nu}_\tau$ unaffected, is an additional 
 tensor operator in the effective $b \to c \ell \nu$ Hamiltonian. Its coupling $\epsilon_T$  has been constrained  using experimental data and employed to compute several  observables. The most sensitive one to discriminate the NP scenario from  SM  is
the forward-backward asymmetry  in  $ B \to D^* \tau {\bar \nu}_\tau$, with a shift in  the position of its zero.
In the case of $B$ decays to excited charmed mesons, the NP model predicts   an enhancement of the ratios  ${\cal R}$   with respect to SM and a sizable modification of the  forward-backward asymmetries   in the decays to $D_1^\prime$ and $D_2^*$.
\Acknowledgements
I thank P. Biancofiore and P. Colangelo for collaboration on the topics discussed here.


\begin{thebibliography}{99}


\bibitem{Lees:2012xj}
J.~P.~Lees {\it et al.}  [BaBar Collaboration],
Phys.\ Rev.\ Lett.\  {\bf 109}, 101802 (2012).

\bibitem{Fajfer:2012vx}
  S.~Fajfer, J.~F.~Kamenik and I.~Nisandzic,
  Phys.\ Rev.\ D {\bf 85}, 094025 (2012).


  \bibitem{altri}
  S.~Fajfer {\it et al.},
  Phys.\ Rev.\ Lett.\  {\bf 109}, 161801 (2012);
  A.~Crivellin  {\it et al.},
  Phys.\ Rev.\ D {\bf 86}, 054014 (2012);
Phys.\ Rev.\ D {\bf 87}, no. 9, 094031 (2013);
  A.~Datta  {\it et al.},
  Phys.\ Rev.\ D {\bf 86}, 034027 (2012);
  D.~Becirevic  {\it et al.},
  Phys.\ Lett.\ B {\bf 716}, 208 (2012);
   A.~Celis {\it et al.},
  JHEP {\bf 1301}, 054 (2013);
  D.~Choudhury  {\it et al.},
  Phys.\ Rev.\ D {\bf 86}, 114037 (2012);
  M.~Tanaka and R.~Watanabe,
  Phys.\ Rev.\ D {\bf 87},  034028 (2013);
  I.~Doršner  {\it et al},
  JHEP {\bf 1311}, 084 (2013);
  Y.~Sakaki  {\it et al},
  Phys.\ Rev.\ D {\bf 88}, no. 9, 094012 (2013);
  M.~Duraisamy {\it et al},
  Phys.\ Rev.\ D {\bf 90}, 074013 (2014).

\bibitem{Adachi:2012mm} 
  I.~Adachi {\it et al.}  [Belle Collaboration],
  Phys.\ Rev.\ Lett.\  {\bf 110}, no. 13, 131801 (2013);
  J.~P.~Lees {\it et al.}  [BaBar Collaboration],
  Phys.\ Rev.\ D {\bf 88}, no. 3, 031102 (2013).



\bibitem{Biancofiore:2013ki} 
  P.~Biancofiore, P.~Colangelo and F.~De Fazio,
  Phys.\ Rev.\ D {\bf 87},  074010 (2013).



\bibitem{Isgur:1989ed}
  N.~Isgur and M.~B.~Wise,
  Phys.\ Lett.\ B {\bf 237}, 527 (1990).


\bibitem{Caprini:1997mu}
  I.~Caprini, L.~Lellouch and M.~Neubert,
  Nucl.\ Phys.\ B {\bf 530}, 153 (1998).

\bibitem{Aubert:2009ac}
  B.~Aubert {\it et al.}  [BABAR Collaboration],
  Phys.\ Rev.\ Lett.\  {\bf 104}, 011802 (2010).


\bibitem{Dungel:2010uk}
  W.~Dungel {\it et al.}  [Belle Collaboration],
  Phys.\ Rev.\ D {\bf 82}, 112007 (2010).

\bibitem{Lees:2013uzd} 
  J.~P.~Lees {\it et al.}  [The BaBar Collaboration],
  Phys.\ Rev.\ D {\bf 88}, 072012 (2013).

\bibitem{Colangelo:2012xi}
 P.~Colangelo {\it et al.},
  Phys.\ Rev.\ D {\bf 86}, 054024 (2012).

\bibitem{Colangelo:1992kc}
  P.~Colangelo, G.~Nardulli and N.~Paver,
  Phys.\ Lett.\ B {\bf 293}, 207 (1992);\\
  P.~Colangelo, F.~De Fazio and N.~Paver,
  Nucl.\ Phys.\ Proc.\ Suppl.\  {\bf 75B}, 83 (1999).
 
\bibitem{Colangelo:1998ga}
  P.~Colangelo, F.~De Fazio and N.~Paver,
  Phys.\ Rev.\ D {\bf 58}, 116005 (1998).


\bibitem{Morenas:1997nk}
  V.~Morenas {\it et al},
  Phys.\ Rev.\ D {\bf 56}, 5668 (1997).

\bibitem{Blossier:2009vy}
  B.~Blossier {\it et al.}  [European Twisted Mass Collab.],
  JHEP {\bf 0906}, 022 (2009).

\end{thebibliography}
\end{document}